\documentclass[a4paper,10pt,twoside]{cpc-hepnp}
\usepackage{CJK,upgreek,fancyhdr}
\usepackage{multicol}
\usepackage{graphicx}
\usepackage{booktabs}
\usepackage{amssymb,bm,mathrsfs,bbm,amscd}
\usepackage[tbtags]{amsmath}
\usepackage{lastpage}

\def\od{{\rm d}}
\def\bc{\begin{center}}
\def\ec{\end{center}}
\def\be{\begin{eqnarray}}
\def\ee{\end{eqnarray}}

\newcommand{\omits}[1]{}

\begin{document}
\begin{CJK*}{GB}{gbsn}

\fancyhead[c]{\small Chinese Physics C~~~Vol. xx, No. x (201x) xxxxxx}
\fancyfoot[C]{\small 010201-\thepage}

\footnotetext[0]{Received 12 August 2016}

\title{Is GW151226 a really signal of gravitational wave?\thanks{Supported by National Natural Science
Foundation of China (11375203, 11275207) }}

\author{%
      Zhe Chang (³£ÕÜ)$^{1)}$\email{changz@ihep.ac.cn}%
\quad Chao-Guang Huang (»Æ³¬¹â)$^{2)}$\email{huangcg@ihep.ac.cn}%
\quad Zhi-Chao Zhao (ÕÔÖ¾³¬)$^{3)}$\email{zhaozc@ihep.ac.cn}
}
\maketitle

\address{%
Institute of High Energy Physics, Chinese Academy of Sciences, Beijing 100049, China\\
School of Physics, University of Chinese Academy of Sciences, Beijing 100049, China
}

\begin{abstract}Recently, the LIGO Scientific Collaboration and Virgo Collaboration published the second observation on gravitational wave GW151226 [Phys. Rev. Lett. {\bf 116}, 241103 (2016)] from the binary black hole coalescence with initial masses about 14 M$_\odot$ and 8 M$_\odot$.  They claimed that the peak gravitational strain was reached at about 450 Hz, the inverse
of which has been longer than the average time a photon staying in the Fabry-Perot cavities in
two arms.  In this case, the phase-difference of a photon in the two arms due to the
propagation of gravitational wave does not always increase as the photon stays in the
cavities.  It might even be cancelled to zero in extreme cases.  When the propagation effect is taken into account, we find that the claimed signal GW151226 would almost disappear.
\end{abstract}



\footnotetext[0]{\hspace*{-3mm}\raisebox{0.3ex}{$\scriptstyle\copyright$}2013
Chinese Physical Society and the Institute of High Energy Physics
of the Chinese Academy of Sciences and the Institute
of Modern Physics of the Chinese Academy of Sciences and IOP Publishing Ltd}%

\begin{multicols}{2}

\section{Introduction}
The analysis of GW150914 shows that the initial black hole masses
are 36M$_\odot$ and 29M$_\odot$ \cite{LVC00}, which are heavier than the previous
known stellar-mass black holes\cite{StellarBHmass}.  In the newly
announced black hole merge event, GW151226 \cite{LVC10}, the initial
black hole masses are about 14M$_\odot$ and 8M$_\odot$, which fall
into the known mass range of stellar black holes in the previous
observations.  It seems to make the picture of binary black hole merge
and gravitational wave observation more reliable
because the signals of GW150914 and GW151226 are
extracted from noise by the same methods\cite{{Black},{MatchFilter}}.

However, we notice the response of a detector to gravitational wave is a function of frequency. When the time a photon moving around in the Fabry-Perot cavities is the same order of the period of a gravitational wave, the phase-difference due to the gravitational wave should be an integral along the path. In fact, this propagation effect on Michelson detector response was addressed, for example, in \cite{cqg}. Unfortunately, the propagation effect on Fabry-Perot detector response has not been considered properly. In the manuscript, we try to take into the propagation effect of the gravitational wave and reexamine the LIGO data. We find that when the average time a photon staying in the Fabry-Perot cavities in two arms is the same order of the period of a gravitational wave, the phase-difference of a photon in the two arms due to the gravitational wave may be cancelled. In the case of observation for GW151226, the average time of a photon staying in the detector is longer than the period of the gravitational wave at maximum gravitational radiation. When the propagation effect is taken into account, the claimed signal GW151226 almost disappears.

It is well known that a Michelson interferometer is a broadband gravitational wave detector. The Michelson interferometer with a 4 km arm can be used to detect gravitational waves from tens Hz to thousands Hz. When an interferometer using Fabry-Perot cavities instead of Michelson interferometer arms, the detection sensitivity is multiplied by a factor of  $2N(1-N\epsilon)$, where $N$ is the average number of photons back and forth in the a cavity and $\epsilon$ is the fractional power lost in one round trip inside the a cavity. The above statement seems that the Fabry-Perot cavities do not affect the frequency range of the detector. In fact, there are subtle differences that should be dealt with carefully. In the following, we give a description on details of the propagation effect of gravitational wave on detectors. A 4 km long Michelson interferometer can be used to detect gravitational waves with frequency from tens to thousands Hz. The seismic noise and environmental disturbances set the lower frequency limit of a detector. The upper limit comes from two aspects. The first, astrophysical gravitational waves with frequency over 10 kHz are too weak to detect\cite{Detector}. The second can be seen clearly from the following discussion.

\section{Photons moving in the Fabry-Perot cavities with gravitational wave as background}
Consider the simplest case, a monochromatic gravitational wave with
plus polarization normally incidents on the detector. Assume the
polarization direction be just consistent with the direction of the
arms. The time difference of light travels to the end mirror and
back in two arms is

\begin{eqnarray}
\Delta {{t}_{\text{M}}} &:=&\frac{2}{c}\left (\oint
(1+\frac{1}{2}h_{11})\od x -\oint
(1-\frac{1}{2}h_{11})\od y\right ) \notag \\
&=&\frac{1}{c}\left(\int_0^{L}h_{11}\od x-\int_{L}^0h_{11}\od x\right .\notag \\
&& \qquad \left .+\int_0^{L}h_{11}\od y -\int^0_{L}h_{11}\od y \right) \notag \\
&=& \frac{4L}{c}\frac{\sin(2\pi f L/c)}{2\pi f L/c} h_{11}(t+L/c)\mid_{z=0}, \label{eq:1rt}
\end{eqnarray}
where $z=0$ denotes the plane of the detector, $f$ is the frequency of gravitational wave, $L$ is the length of the detector's arm, $c$ is the speed of light, and $h_{11}(t-L / c)$ is the gravitational wave at the time $t-L/c$. Thus, the laser field acquires a phase shift as it travels to the end mirror and back

\begin{eqnarray}
\Delta {{\Phi }_{\text{M}}}&=&2\pi \nu \Delta {{t}_{\text{M}}} \notag\\
&=&\frac{2L}{\lambda /2\pi }\frac{\sin (2\pi fL/c)}{2\pi fL/c}{{h}_{11}}(t-L/c){{}_{z=0}}.
\end{eqnarray}
At the case of $fL/c=1/2$ (which means $f=c/(2L)=37500$Hz ), the signal is cancelled just.
That is to say, the detection frequency of the Michelson interferometer for gravitational wave shall not exceed $c/(4L)=18750$Hz . Similar result was addressed in \cite{cqg}. When the frequency of gravitational wave is $1$ kHz or less, the time difference and phase shift can be written as

\begin{eqnarray}
\Delta {{t}_{\text{M}}}=\frac{2L}{c}{{h}_{11}}(t+L/c){{}_{z=0}},
\end{eqnarray}

\begin{eqnarray}
\Delta {{\Phi }_{\text{M}}}=2\pi \nu \Delta {{t}_{\text{M}}}=\frac{2L}{\lambda /2\pi }{{h}_{11}}(t+L/c){{}_{z=0}}.
\end{eqnarray}

To increase the response of detector to gravitational wave, the Fabry-Perot resonant cavities are used to make photons back and forth for many times. Denote the average number of round trips of photons in the arm as $N$ . When $N$ is small, we can still use formulae (3) and (4). It is a good approximation. However, if $N$ is large enough, we should use the general expressions (1) and (2). For $1$ kHz gravitational wave, this approximation is no longer hold yet if $N>40$ . It should be noticed that for the advanced LIGO, $N = 140$. So we should reconsider the approximation.

The above analysis seems to be contrary to the fact that a Fabry-Perot cavity can improve the detection sensitivity multiples. Take into account the introduction of sidebands, Schnupp asymmetry, and PDH locking technique, the response of Michelson interferometer is

\begin{eqnarray}
{{V}_{s}}={{P}_{0}}R{{J}_{0}}\left( \beta  \right){{J}_{1}}\left( \beta  \right) kLh{{\sin }^{2}}\frac{2\pi \Delta L}{{{\lambda }_{\Omega }}},
\end{eqnarray}
where $P_{0}$  is the power of incident laser, $R$  is the response of the photodiode, $J_{0}$ and $J_{1}$ are the zeroth and first order Bessel function, $\beta$ is the phase modulation parameter, $k$ is the wave number of carrier wave of the detector, $\Delta L$ is the length difference of the two arms, $\lambda_{\Omega}$ is the wavelength of sideband\cite{SIDEBAND}. For the Fabry-Perot cavities, the response of gravitational wave becomes as
\begin{eqnarray}
{{V}_{s}}={{P}_{0}}R{{J}_{0}}\left( \beta  \right){{J}_{1}}\left( \beta  \right)\frac{2F}{\pi }\left( 1-\frac{F}{\pi }\varepsilon  \right)\sin \left( 2\pi \frac{\Delta L}{{{\lambda }_{\Omega }}} \right)kLh,
\end{eqnarray}
where $F/\pi=N$. Eqs. (5) and (6) show that the using of the Fabry-Perot cavities increases response by a factor of $2N$ when the loss is ignored. However, Eq.(6) is derived in the assumption of low frequency limit (i.e. $2NL << c/f$ ). Our analysis on GW151226 based on the light moving in the cavities, is not contrary to the property of Fabry-Perot cavities.


Now, let's consider Eqs. (2) (3) and (4) for LIGO detectors.
For LIGO detectors, the lengths of Fabry-Perot cavities are $L \approx 4$ km.
On average, a photon travels in cavities 140 round trips\cite{LVC00}.  It will move back
and forth in cavities for about 0.0037 s.  In that period, the gravitational wave with frequency
268 Hz has propagated the distance of one wavelength.  Therefore, the above propagation
effect should be taken into account in the analysis of GW151226 because the frequency of
the peak gravitational strain is about 450 Hz ($>268$ Hz).

\section{LIGO-GW151226 gravitational wave signal revisited}

LIGO Scientific Collaboration provides a simple program which can be used to search
gravitational wave signals from the noise data. They also provide a set of strain data
for two detectors and a well matched template.  In the program, the various subtleties,
whose effects are very small, are ignored on LIGO website\cite{LIGOWebsite}.
With the detector parameters, the program and the noise data, the propagation effect of
gravitational wave on the signal GW151226 can be re-examined.

\begin{center}
\includegraphics[width=8cm]{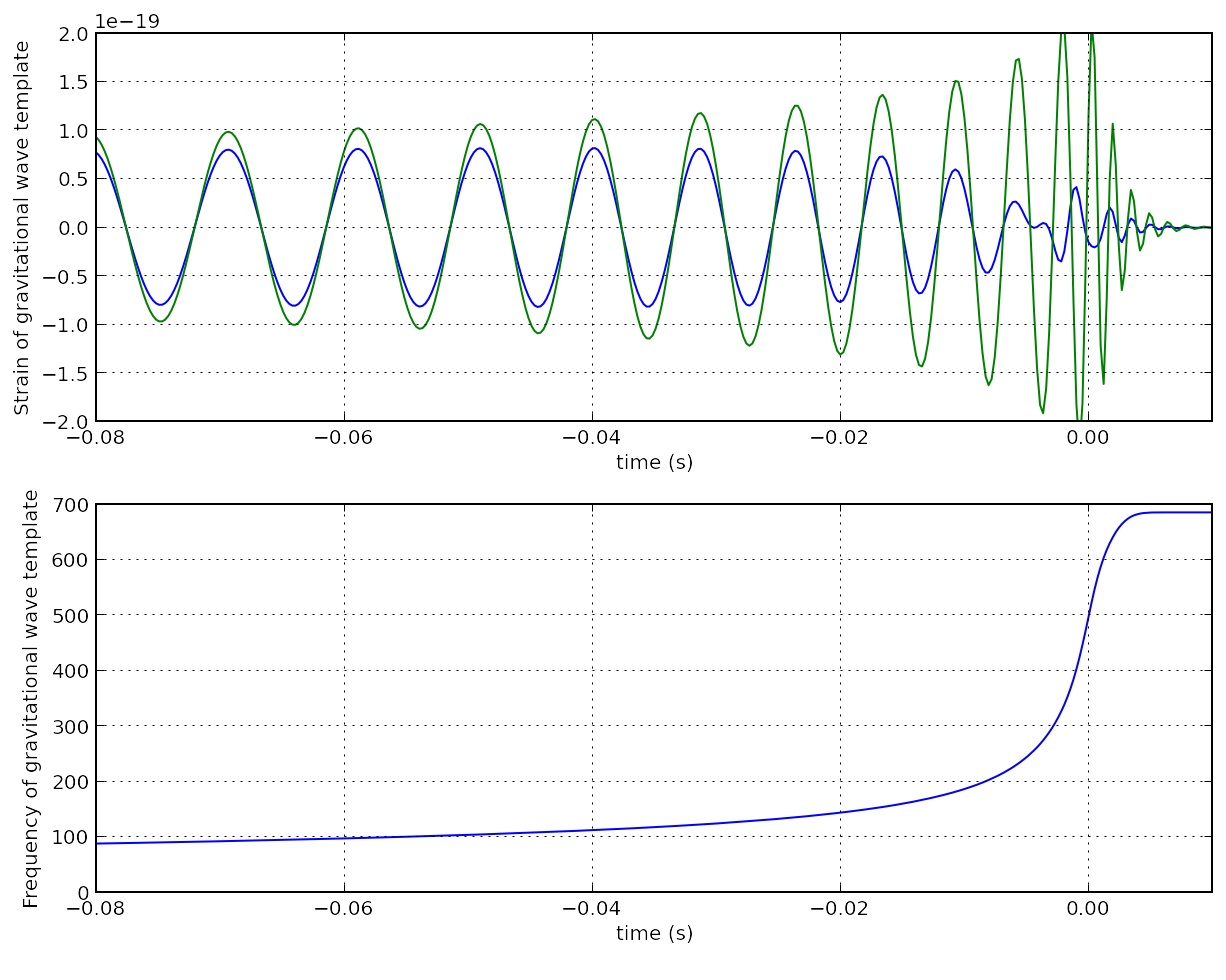}
\figcaption{\label{fig1}   The green line in the top panel is the response of detector to the best fit template for GW151226
provided on LIGO website\cite{LIGOWebsite}.When the propagation effect is taken into account, the detector
response to the gravitational wave in the template becomes the form of the blue line.The bottom panel
 presents the variation of frequency in time for gravitational wave. }

\end{center}

\begin{center}
\includegraphics[width=8cm]{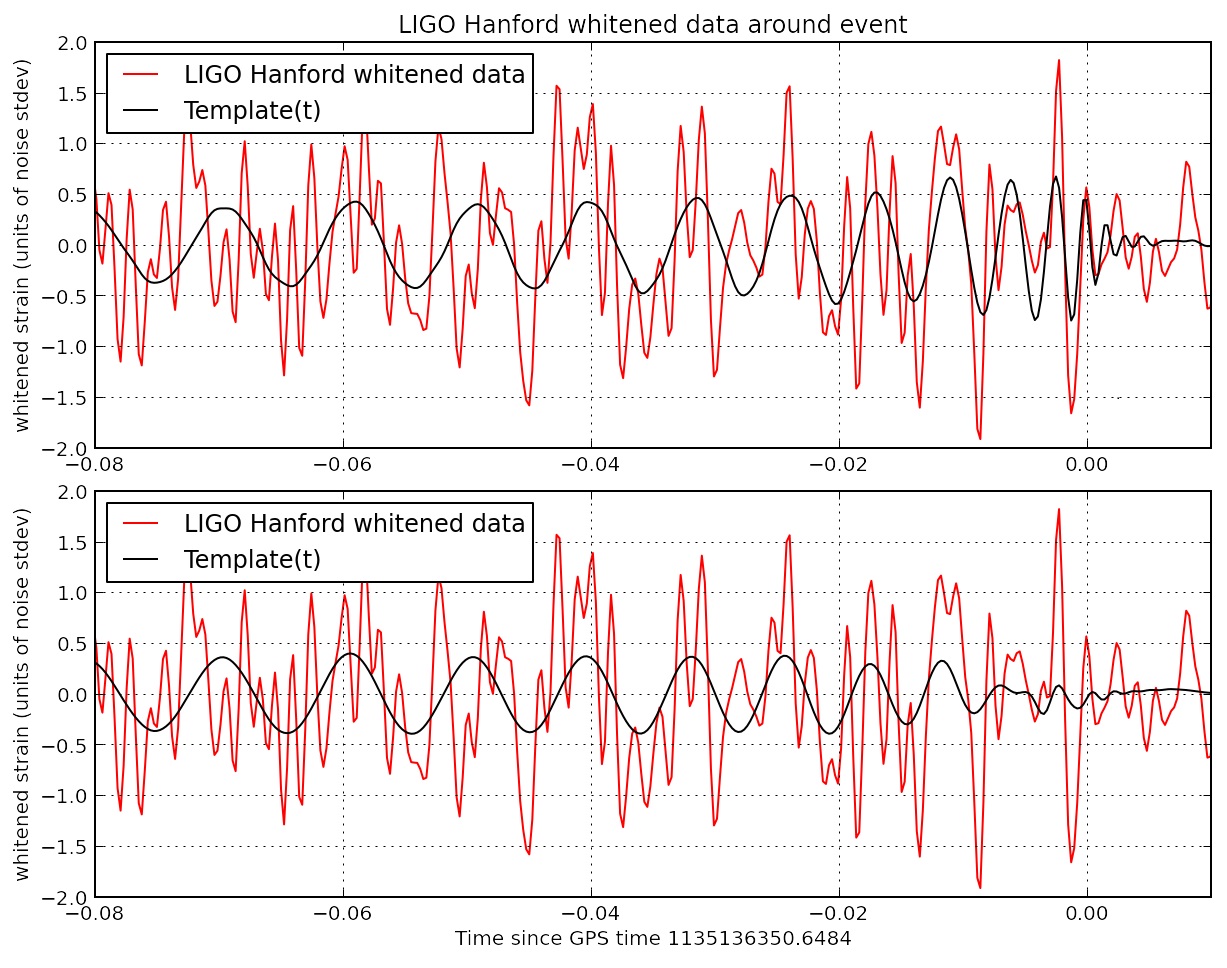}
\figcaption{\label{fig1}   The upper panel of Fig. 2 is the match of the data from the detector at Hanford and the
response to the gravitational wave in the template provided on LIGO website.  The lower panel shows that
the data from the detector does not match the response of detector to the gravitational wave in the template if the propagation effect of gravitational wave is taken into account. }
\end{center}

The green line in the top panel of Fig. 1 is the response of detector to the best fit template for GW151226
provided on LIGO website\cite{LIGOWebsite}.  Near the instance the coalescence happens,
the signal is strongest.  When the propagation effect is taken into account, the detector
response to the gravitational wave in the template becomes the of the blue line.  It
shows that the response of gravitational wave on detector becomes tiny when the real time
of a cycle of gravitational wave is near the time a photon travels in arms.  The bottom panel
of Fig. 1 presents the variation of frequency in time for gravitational wave.  The vanishing response happens at
about 268 Hz, at the case the time of a cycle of gravitation wave is 0.0037 s.
The upper panel of Fig. 2 is the match of the data from the detector at Hanford and the
response to the gravitational wave in the template provided on LIGO website.  The lower panel of Fig. 2 shows that
the data from the detector does not match the response of detector to the gravitational wave in the template if the propagation effect of gravitational wave is taken into account.
{It should be noted that in FIGs. 1 and 2 the templates are obtained by taking into account of the propagation effect  with varying frequency and amplitude,  which is more accurate than what shown at \cite{CS}.}

\section{Conclusion and remark  }

As a conclusion, the propagation effect of gravitational wave is
important in matching the signals in observations with templates
when the time of a cycle of gravitational wave is the same order as
the time a photon stays in cavities.  By taking into account of the
propagation effect of gravitational wave, we find that the
LIGO-GW151226 signal is almost disappear. For the low-frequency
gravitational wave the propagation effect is small. So the signal
for GW150914 is not affected a lot.

It should be remarked that there is the subtle difference between the effect of a gravitational wave on the light traveling in a detector and the phase variation due to the vibration of mirrors which has been used in the calibration of LIGO's detectors\cite{clb} though both the vibration of mirrors and the incidence of a gravitational wave will modify the phase of a light travels in the cavities. The vibrations of the mirrors modify the phase of a light when the photons travel near the vibrating mirrors. The phase shift of a light beyond the vibration region will not be affected by the vibrating mirrors. In contrast, a gravitational wave affects the phase of a light at every place in the cavities. As the result, the average phase variations due to the vibrating mirrors do not vanish even when the time of a round trip of a photon in a cavity is the same as the period of the vibration of the end mirrors. But, it will vanish in the gravitational wave background when the time of a round trip for a photon is the same as the period of a gravitational wave.  Therefore, the propagation effect of gravitational wave is not included in the calibration, which is calibrated with the help of the vibrating mirrors.

\end{multicols}

\clearpage
\end{CJK*}

\bigskip

\begin{thebibliography}{90}

\vspace{3mm}
%
%
%

\bibitem{LVC00} B. P. Abbott, et al,
Phys. Rev. Lett. {\bf 116}: 061102 (2016)

\bibitem{StellarBHmass} J. A. Orosz, et al,
Nature 449: 872--875 (2007)

\bibitem{LVC10}B. P. Abbott, et al,
Phys. Rev. Lett. {\bf 116}: 241103 (2016)

\bibitem{Black} For example, see: E. D. Black and R. N. Gutenkunst,
Am. J. Phys. 71: 365--378 (2003)

\bibitem{MatchFilter} B. Allen, W. G. Anderson, P. R. Brady, D. A. Brown, and J. D. E. Creighton,
Phys. Rev. D 85: 122006 (2012)


\bibitem{cqg} Rakhmanov, M., J. D. Romano, and John T. Whelan,
Class.Quantum Grav. 25.18 (2008): 184017.

\bibitem{Detector} K. S. Thorne,
in \emph{Three hundred years of gravitation}, edited by S. W. Hawking and W. Israel,
(Cambridge University Press, Cambridge, 1987), p.330

\bibitem{SIDEBAND} E.D. Black and R.N.Gutenkunst,
Am. J. Phys. 69: 79-87 (2001)

\bibitem{LIGOWebsite} https:$\backslash\backslash$losc.ligo.org$\backslash$tutorials$\backslash$

\bibitem{CS} Chang Z, Huang CG, Zhao ZC,
Sci. China--Phys. Mech. Astro. 59 100421 (2016)

\bibitem{clb} Abbott, B.P. and LIGO Scientific Collaboration,
arXiv:1602.03845, 2016

\end{thebibliography}
\end{document}